\documentclass[fleqn,10pt]{wlscirep}
\usepackage[utf8]{inputenc}
\usepackage[T1]{fontenc}
\usepackage{algorithmic}
\usepackage[ruled]{algorithm2e}
\usepackage{bm}
\usepackage{chngcntr}
\usepackage{natbib}
\usepackage{multirow}
\usepackage{pifont}
\usepackage{makecell}
\setcitestyle{numbers}

\title{RapidMitDet: Boosting Computational Efficiency for Mitosis Detection via Magnification Alignment}
	
	\author[1,2,3,$\dagger$]{Longfei Wang}
	\author[2,3,4,$\dagger$]{Jiatai Lin}
	\author[1,2,3]{Junyu Guo}
	\author[1,2,3]{Chaoqun Chen}
	\author[5,6]{Danyi Li}
	\author[7]{Lili Zhai}
	\author[1,2,3]{Jingqi Huang}
	\author[2,3]{Tianpeng Deng}
	\author[2,3]{Zhenwei Shi}
	\author[8]{Ying Wang}
	\author[5,6]{Changhong Liang}
	\author[7]{Xiaojing Guo}
	\author[2,3]{Li Liang}
	\author[2,3,*]{Zaiyi Liu}
	\author[2,3,*]{Chu Han}
	
	\affil[1]{School of Biomedical Engineering, Southern Medical University,1023 Shatai Rd., Baiyun District,Guangzhou,510515,Guangdong,China}
	\affil[2]{Department of Radiology, Guangdong Provincial People's Hospital (Guangdong Academy of Medical Sciences), Southern Medical University,Guangzhou,510080,Guangdong,China}
	\affil[3]{Guangdong Provincial Key Laboratory of Artificial Intelligence in Medical Image Analysis and Application,Guangzhou,510080,Guangdong,China}
	\affil[4]{Guangdong Cardiovascular Institute, Guangdong Provincial People's Hospital, Guangdong Academy of Sciences,Guangzhou,510080,Guangdong,China}
	\affil[5]{Department of Pathology, Nanfang Hospital and Basic Medical College, Southern Medical University,Guangzhou,510515,Guangdong,China}
	\affil[6]{Guangdong Province Key Laboratory of Molecular Tumor Pathology,Guangzhou,510515,Guangdong,China}
	\affil[7]{Department of Breast Pathology and Laboratory, Tianjin Medical University Cancer Institute and Hospital, National Clinical Research Center for Cancer, Key Laboratory of Breast Cancer Prevention and Therapy of Ministry of Education of China, Tianjin Medical University, Tianjin's Clinical Research Center for Cancer,West Huanhu Road,Tianjin,300060,China}
	\affil[8]{Department of Medical Ultrasonics, The First Affiliated Hospital of Guangzhou Medical University,151 Yanjiang West Road,Guangzhou,510120,Guangdong,China} 
	
\affil[$\dagger$]{These authors contributed equally to this work.}
\affil[*]{Correspondence:  {liuzaiyi@gdph.org.cn} (Zaiyi Liu), {hanchu@gdph.org.cn} (Chu Han) }
	
	\begin{abstract}
		Automated mitosis detection is a critical research focus, enabling pathologists to rapidly assess mitotic indices for tumor malignancy evaluation. However, most of the existing detection approaches suffer from significant computational inefficiency due to the giga-pixel image resolution of Whole-Slide Images (WSIs), which greatly hinders their clinical applicability. In this study, we propose RapidMitDet, a computational efficient mitosis detection approach. Our method employs a coarse-to-fine framework, first identifying mitotic candidates at lower magnification and then confirming them at higher magnification. In the coarse-detection stage, we introduce magnification-aligned knowledge distillation to reduce the information gap between magnification levels while minimizing the performance degradation in the lower magnification detection model. In the subsequent fine-classification stage, we propose adaptive triplet learning (ATL) to optimize infomative sample selection, thereby enhancing discrimination between mitotic and non-mitotic instances. Experimental results show that RapidMitDet achieves the optimal accuracy-efficiency balance at 20$\times$ magnification, improving computational speed by up to 5.62 times over the latest state-of-the-art (SOTA) methods. Due to its lightweight model design, RapidMitDet can also deliver efficient detection at higher magnification (40$\times$) while achieving even higher detection performance. In WSI-level evaluations, our method shows strong agreement with pathologists' diagnoses (ICCs: 0.879 (40$\times$), 0.822 (20$\times$)) while maintaining superior efficiency, highlighting its potential for clinical application. Code is available at \href{https://github.com/logincat/RapidMitDet}{https://github.com/logincat/RapidMitDet}
	\end{abstract}
	
\begin{document}
	
	\flushbottom
	\maketitle
	
	\thispagestyle{empty}

\section*{KEYWORDS}
Computational Pathology, Mitosis Detection, Computational Efficiency, Magnification Alignment

\section{Introduction}
Mitotic count serves as a critical biomarker of cellular proliferation~\citep{collan1991value}, with a well-established prognostic value in major cancer grading systems, including breast (Nottingham) and soft tissue sarcoma (FNCLCC)~\citep{wang2022improved, schwen20201659p}. However, its clinical practice is hampered by the needle-in-a-haystack nature of mitosis (<0.1\% cell prevalence), introducing significant inter-observer variability and workflow inefficiencies~\citep{aubreville2023comprehensive}.

\begin{figure}[tb]
	\centering
	\includegraphics[width=\linewidth]{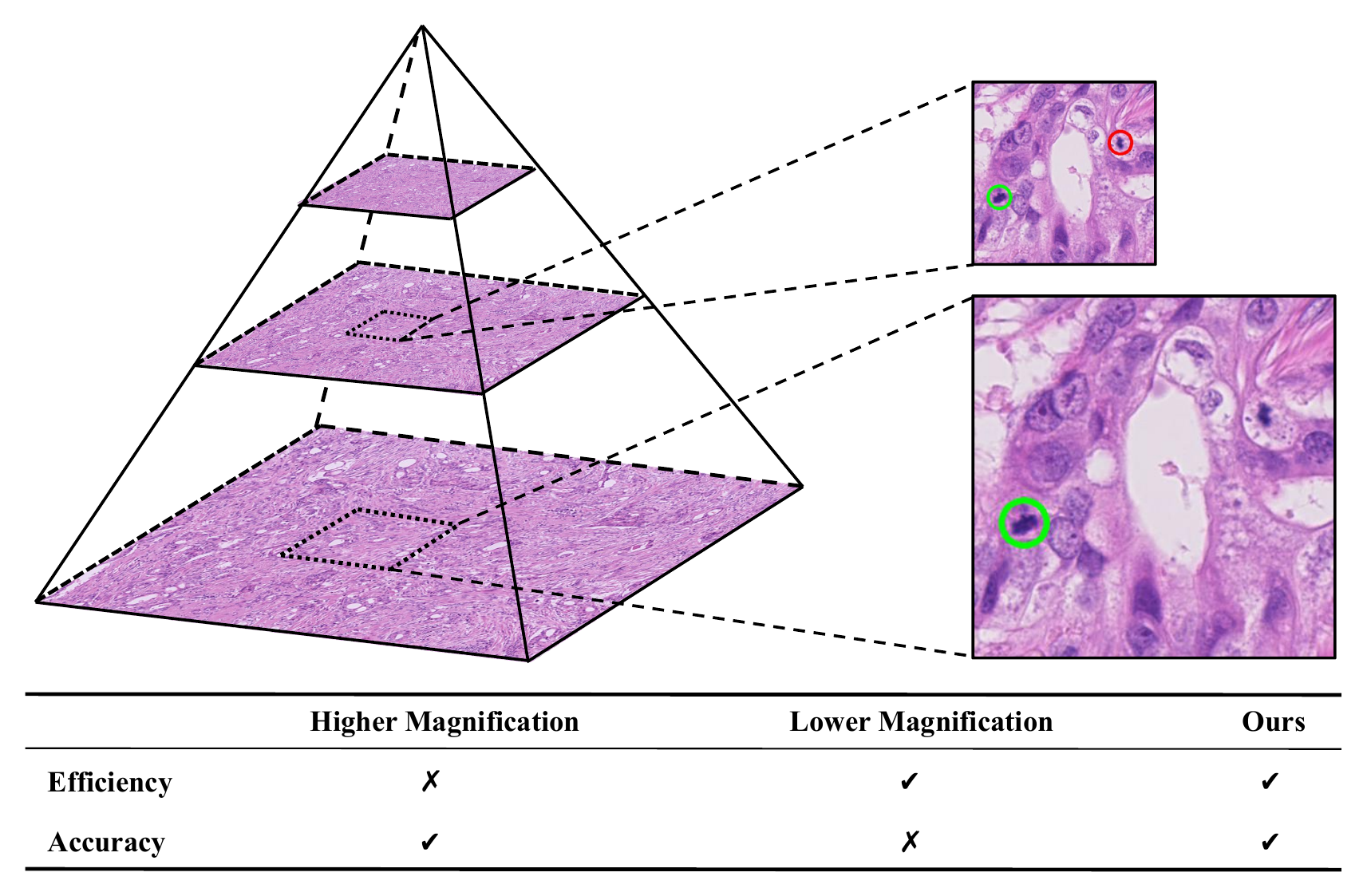}
	\caption{The effect of different magnifications on accuracy and efficiency. Color of circle border~(green: true positive, red: false positive).}
	\label{fig1}
\end{figure}

Recent advances in computational pathology have sought to automate mitotic count through deep learning. Current approaches achieve accurate mitosis detection by addressing key challenges such as class imbalance, domain shift, and etc~\citep{lin2025rethinking,wang2023generalizable}, which relies on highest-magnification inference. While highest-magnification processing maintains high detection performance, the computational overhead imposed by gigapixel whole-slide images inevitably hinders real-world clinical deployment as shown in Fig.~\ref{fig1}.

The growing demand for efficient computational pathology has made optimization of processing speed a crucial research priority~\citep{van2021deep}. It is particularly evident in mitosis detection, where current fully automated approaches often face a fundamental trade-off between computational efficiency and detection accuracy~\citep{jahanifar2024mitosis}. Clinically, pathologists routinely employ a hierarchical strategy to improve mitosis detection efficiency by rapid screening at lower magnification and detailed confirmation at higher magnification.

Inspired by the above clinical knowledge, we propose a novel feasible mitosis detection pipeline to achieve this goal, named \textbf{RapidMitDet}, which conducts a coarse screening at lower magnification and then achieves fine identification at higher magnification. To reconcile efficiency with mitosis detection performance, the coarse stage of our RapidMitDet guides the lower magnification detection model to learn the feature representation of higher magnification by magnification alignment. Then, the fine classification stage employs the triplet learning for higher-magnification model training. To improve the classification accuracy, we proposed adaptive triplet learning~(ATL) by selecting the nearest negative sample of the anchor and maximizing the distance between them, thus further expanding the inter-cluster differences in feature representation.

Finally, we conduct massive experiments to demonstrate the effectiveness and efficiency of the proposed RapidMitDet. In the experiments, our RapidMitDet has achieved state-of-the-art~(SOTA) accuracy and efficiency on three public mitosis detection datasets. In external validation using clinical whole slide images~(WSIs), RapidMitDet has demonstrated superior diagnostic concordance (ICC=0.879) with pathologists while achieving 8.07$\times$ faster processing speeds versus existing methods. In summary, the main contributions of this paper include:
\begin{itemize}
	\item We propose a magnification alignment (MAG) module that effectively aligns feature representations across different magnifications within a coarse-to-fine mitosis detection framework, improving both detection efficiency and robustness.
	\item We propose an adaptive triplet learning (ATL) module that dynamically constructs informative triplets through adaptive opposite-class sample selection, resulting in a more discriminative feature representation for fine mitosis classification.
	\item In the experiments, RapidMitDet achieves state-of-the-art performance on three public datasets while demonstrating strong clinical consistency with pathologists and faster inference on in-house whole slide images~(WSIs) compared with existing methods.
\end{itemize}

\section{Related Works} \label{related work}
\subsection{Mitosis detection}
Mitosis detection is a fundamental task of computational pathology. It plays a pivotal role in quantifying pathological indicators related to mitosis~\citep{cree2021counting}. In recent years, numerous public mitosis datasets have driven the progress of mitosis detection, such as MIDOG2021~\citep{aubreville2023mitosis}, MIDOG2022~\citep{aubreville2024domain}, MIDOG++~\citep{aubreville2023comprehensive}. Meanwhile, recent studies have developed numerous methods, which can be classified methodologically into segmentation-based, detection-based, and classification-based approaches.

\textbf{Classification-based approaches:} These approaches usually use a sliding-window technique to extract patches and classify whether they contain mitotic figures. Despite their success, classification-based approaches inherently lack pixel-level localization, thereby severely limiting their precision in pinpointing the locations of mitotic cells. To address this issue, \cite{pati2020mitosis} employs an additional localization tool, such as CAM~\citep{zhou2016learning}, to ascertain the locations of mitotic cell centers. However, filtering all patches is undoubtedly inefficient. \cite{lin2025rethinking} employs a hematoxylin-based method to identify candidate mitotic cell centers, which effectively reduces unnecessary classification, thereby improving the efficiency of the algorithm.

\textbf{Segmentation-based approaches:} These approaches usually transform the mitosis detection task into a segmentation task to achieve precise mitotic localization. To bypass labor-intensive pixel-level annotations, \cite{jahanifar2024mitosis} converts weak box/point labels into masks for training a mitosis segmentation model. However, this coarse conversion method fails to yield precise pixel-level annotations. To address this limitation, \cite{wang2023generalizable} refines the annotation conversion process by deriving accurate mitotic masks through the intersection of HoverNet~\citep{graham2019hover} instance segmentations and bounding boxes, while concurrently enhancing domain generalization via Fourier-based data augmentation.

\textbf{Detection-based approaches:} These approaches usually locate mitotic cells through bounding-box regression. However, the high morphological similarity between mitotic and non-mitotic cells causes detection algorithms to misclassify many non-mitotic cells as mitotic, significantly degrading detection performance. To address this issue, \cite{tang2024deep} constructs a deep radial basis function~(RBF) network with subcategorization to improve feature distinguishability within the subcategory space.

Mitosis detection research now prioritizes computational efficiency alongside accuracy. \cite{jahanifar2024mitosis} proposed a fast-and-slow paradigm~(MDFS) to achieve efficient mitosis detection through lower-magnification screening followed by higher-magnification precise classification, which offers valuable insights. However, MDFS did not specifically address the challenge of performance degradation at lower magnifications. Therefore, our work aims to resolve the efficiency–accuracy trade-off in the coarse-to-fine pipeline.

\subsection{Knowledge distillation}
Knowledge distillation transfers potential knowledge from a computationally intensive teacher model to a compact student model, enabling the latter to achieve comparable performance with significantly reduced inference latency.

\textbf{Probability distribution alignment:}
This approach uses softened softmax outputs (logits) from the teacher to guide the student via Kullback–Leibler divergence~(KLD)~\citep{kullback1951information} minimization. To address the limitations of current regression loss designs for high-accuracy detection, \citep{yang2021learning} calculates the KLD between Gaussian distributions as the regression loss, dynamically adjusting the parameter gradients according to the characteristics of the object. However, KLD assumes shared support and may perform poorly when output spaces differ or distributions are disjoint. To mitigate these issues, \citep{yang2021improved} employs the Wasserstein Distance (WD) for knowledge transfer, providing a more geometrically meaningful measure of the distance between probability distributions.

This technique explicitly aligns feature response magnitudes between the teacher and student across layers, ensuring that the student mimics both the teacher's patterns and feature scales. To address the issue of excessive time consumption in pathological image transmission and inference, \cite{han2025towards} employs self-supervised learning to bridge the information gap between low- and high-magnification levels by effectively aligning their feature representations, achieving remarkable efficiency gains.

In this paper, we aim to address performance degradation at lower magnifications and achieve an efficiency–accuracy trade-off by leveraging magnification alignment.

\section{Methodology} \label{method}
This paper proposes a novel strategy, named RapidMitDet, to improve the computational efficiency of mitosis detection. In this section, we first provide an overview of the proposed RapidMitDet in Section~\ref{3.1}. Then, we detail the coarse- and fine-stage mitosis detection processes during the training phase in Sections~\ref{3.2} and~\ref{3.3}, respectively. Finally, we describe the implementation details in Section~\ref{3.4}.

\begin{figure*}[th]
	\centering
	\includegraphics[width=\linewidth]{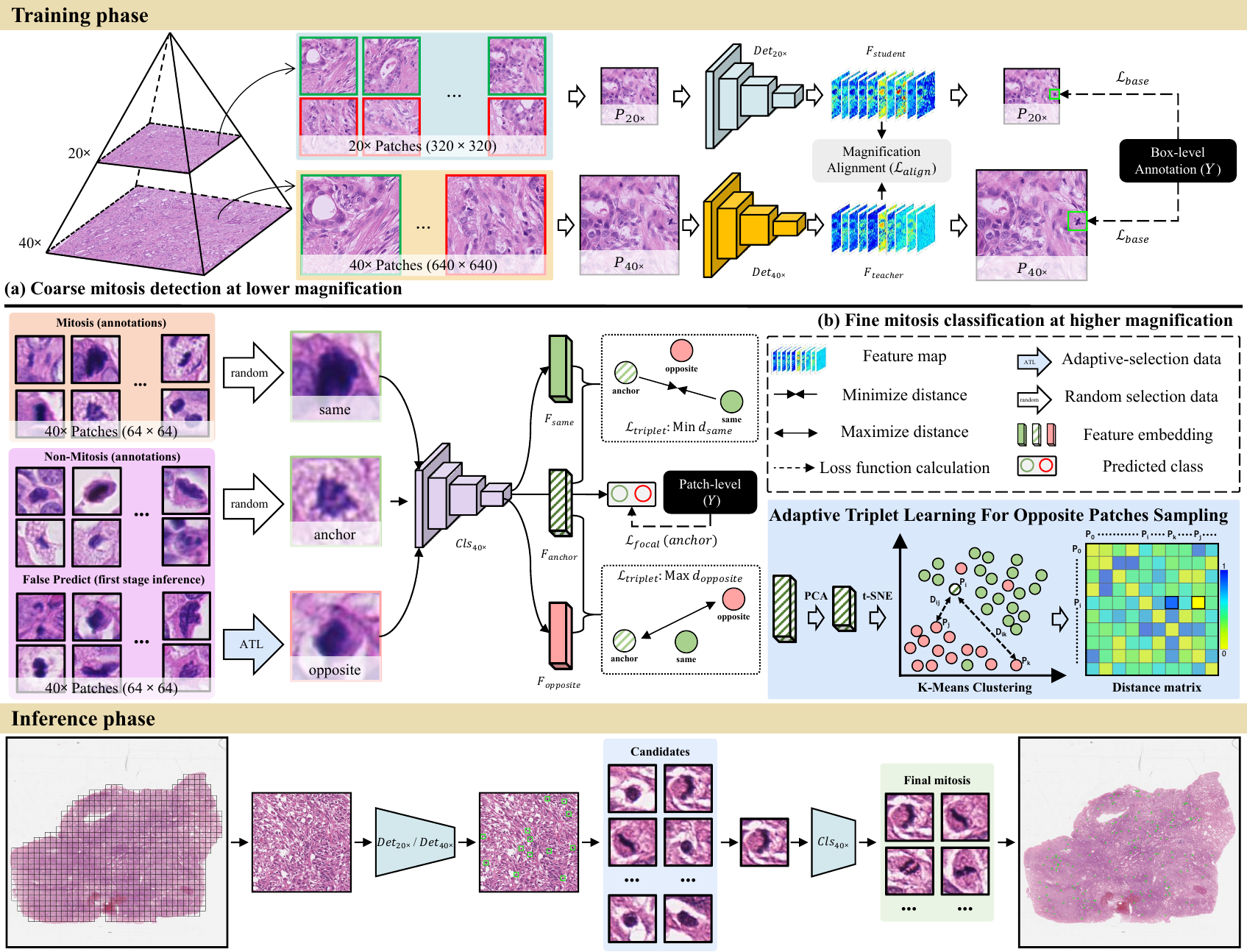} 
	\caption{Framework of the proposed RapidMitDet. Training phase: a) Coarse mitosis detection stage: learning with magnification alignment to train the mitosis detection model at lower magnification. b) Fine mitosis detection stage: mitosis classification model training with the proposed adaptive triplet learning~(ATL) at higher magnification. Inference phase: the corse-to-fine inference pipeline with the well-trained~(20$\times$ or 40$\times$ magnification) models for candidates mitosis detection of whole slide images~(WSIs) and then the well-trained~(40$\times$ magnification) model for further precise classification.} 
	\label{fig2} 
\end{figure*}

\subsection{Overview of our RapidMitDet} \label{3.1}
This paper proposes a coarse-to-fine pipeline for rapid mitosis detection, as shown in Fig.~\ref{fig2}, which performs rapid candidate detection at lower magnification, followed by precise classification at higher magnification.

\textbf{Training phase:}~In our pipeline, we first train the lower-magnification detection model in the coarse stage and then train the higher-magnification classification model in the fine stage.

In the coarse-detection stage,  we employ magnification alignment to train the lower-magnification detection model, enabling it to learn discriminative feature representations from higher magnification and reducing the performance gap between lower- and higher-magnification models. Given the raw training set~$\{\mathcal{X},\mathcal{Y}\}$, where $\mathcal{X}$ represents the histopathological image with pyramid structure and $\mathcal{Y}$ is the corresponding box-level annotation. We use it to create lower- and higher-magnification training sets, denoted $\{X^{det}_{20\times},Y^{det}_{20\times}\}$ and $\{X^{det}_{40\times},Y^{det}_{40\times}\}$. Then, we build the $20\times$ and $40\times$ detection models, denoted ${Det}_{20\times}$ and ${Det}_{40\times}$, which are trained through magnification alignment guided by ${Det}_{40\times}$  at lower magnification and trained by box-level supervision at higher magnification.

In the fine classification stage, we construct a mitosis classification training set at 40× magnification from the raw dataset$\{X^{cls}_{40\times},Y^{cls}_{40\times}\}$, which combines false positive samples generated by the detection model~($Det_{40\times}$), annotated mitosis patches, and annotated hard negative ones. Concurrently, we propose an adaptive triplet learning~(ATL) to improve classification accuracy by maximizing inter-cluster distances in feature space.

\textbf{Inference phase:}~Given an input WSI $\mathcal{X}_{test}$, we employ the detection model~(${Det}_{20\times}$) to screen the candidates of mitosis at lower magnification~($\mathcal{X}^{20\times}_{test}$) by sliding window. The candidates are then cropped as classification patches according to the corresponding coordinates at $40\times$ magnification for further precise classification using $40\times$ classification model $Cls_{40\times}$.

\subsection{Coarse Mitosis Detection with Magnification Alignment}~\label{3.2}
This section is to train a well-detected lower magnification detection model through magnification alignment as shown in Fig.~\ref{fig2}~(a), which contains multi-scale mitosis detection dataset construction, magnification alignment and base supervision.

\textbf{Mitosis detection dataset construction}: Initially, we construct patch-level training sets at two different magnifications (20$\times$ and 40$\times$) to train our detection models through magnification alignment. For the 40$\times$ magnification level, we extract patches of 640$\times$640 pixels, while for the 20$\times$ magnification, we use smaller patches of 320$\times$320 pixels to cover identical tissue regions while providing different levels of resolution for magnification alignment, both cropped from public datasets. We use the box-level annotations provided in the raw dataset, each cropped patch is assigned both an image-level classification label and a box-level detection label.

\textbf{Magnification alignment}: In this step, we align the features between 20$\times$ and 40$\times$ patches that capture identical tissue regions. Adopting the teacher-student knowledge distillation, we leverage the higher magnification 40$\times$ model (teacher) to guide the lower magnification 20$\times$ model (student) to learn the higher magnification feature representations. Both magnification-specific detectors (denoted as ${Det}_{40\times}$ and ${Det}_{20\times}$) employ Hyper-YOLO~\citep{feng2024hyper} as their backbone architecture. Given paired multi-magnification patches, we align their extracted features ($F_{teacher}$ from 40$\times$ and $F_{student}$ from 20$\times$) by the Mean Absolute Error (MAE) loss function:
\begin{equation}
	\mathcal{L}_{align} = {\mathcal{L}_{MAE}}(F_{teacher}, F_{student}) = \frac{1}{N} \sum_{i=1}^{N} \left| F_{teacher}^{(i)} - F_{student}^{(i)} \right|
\end{equation}

where ${L_{MAE}}$ is mean absolute error~(MAE), $F_{teacher}$ and $F_{student}$ are the feature extract by model, consider $F_{teacher}$ as the true value and $F_{student}$ as the predicted value. The ${N}$ is the batch size. 

\textbf{Base supervision:} Subsequently, we employ box-level annotations to train the detection models under box-level supervision, enabling a dense localization of mitotic candidates. In this study, we adopt Mean Squared Error~(MSE) $\mathcal{L}_{MSE}$ to quantify the discrepancy between predicted bounding boxes and ground-truth bounding boxes as follows:
\begin{equation}
	\mathcal{L}_{base} = \mathcal{L}_{MSE}(p_i, y_i) = \frac{1}{N} \sum_{i=1}^N \|  p_i -  y_i \|_2^2
\end{equation}
where $p_i$ and $y_i$ are the $i$-th predicted bounding box and ground truth. $N$ represents the number of samples.

In this stage, we first train the teacher model and then train the student model. When training the teacher model, we only use $\mathcal{L}_{base}$, and when training the student model, we use the total loss function of the coarse detection stage $\mathcal{L}_{coarse}$ as follows:
\begin{equation}
	\mathcal{L}_{coarse} = \mathcal{L}_{base} + \mathcal{L}_{align}
\end{equation}
where $\mathcal{L}_{base}$ and $\mathcal{L}_{align}$ are assigned equal weights.

\subsection{Fine Mitosis Classification with Adaptive Triplet Learning}~\label{3.3}

For fine mitosis classification, we introduce Adaptive Triplet Learning (ATL) to improve feature discriminability by reducing intra-class variation and increasing inter-class separability, as illustrated in Fig.~\ref{fig2}(b). This stage consists of dataset construction, adaptive triplet learning, and base classification supervision.

\textbf{Mitosis classification dataset construction:}
The classification dataset contains annotated mitotic samples, annotated non-mitotic samples, and false-positive samples generated by the first-stage detector. These false-positive samples are incorporated as challenging negative examples to improve the discriminative capability of the classifier. All patches are cropped at 40$\times$ magnification according to annotations or detected coordinates.

Due to the inherent difficulty of exhaustive annotation in medical images, public datasets may contain incomplete annotations, and detector-generated false positives may include potential mislabeled negatives containing true mitotic figures. To investigate the impact of this potential noise, we performed an ablation study by comparing classification performance with and without detector-generated false positives, as shown in Table~\ref{ab_fp}. The results suggest that incorporating hard negative samples is beneficial for classification learning and does not introduce significant degradation under potential annotation noise.

\textbf{Adaptive triplet learning:}
Traditional triplet learning consists of an anchor, a positive sample, and a negative sample. In our implementation, the anchor is randomly selected from either the mitotic or non-mitotic category. The positive sample is randomly selected from the \textbf{same category} as the anchor, while the negative sample is selected from the \textbf{opposite category} using Adaptive triplet learning (ATL).

Given the extracted embeddings $F_{anchor}$, $F_{same}$, and $F_{opposite}$, the triplet loss is defined as:

\begin{equation}
	\mathcal{L}_{triplet}=max(0,d_{as}-d_{ao}+\alpha)
\end{equation}

where $d_{as}$ and $d_{ao}$ represent the Euclidean distances between anchor-same and anchor-opposite embeddings, respectively. The margin $\alpha$ controls the minimum separation between same and opposite samples.

During the first 30 epochs, random triplet sampling is adopted as a warm-up stage to establish an initial feature representation. After this stage, feature embeddings are extracted from the current classifier and projected using PCA ($n\_components=50$). The PCA-reduced features are divided into two clusters using K-Means, and a pairwise Euclidean distance matrix is then computed in the PCA-reduced feature space. t-SNE ($n\_components=2$) is applied to the PCA-reduced features solely for visualization of the clustering results. For each anchor, opposite-class candidates are restricted to samples assigned to the cluster opposite to that of the anchor and further filtered based on their ground-truth (GT) labels to retain only samples from the opposite class. For example, if an anchor is assigned to Cluster 0, candidate samples are drawn from Cluster 1, and only those with the anchor-opposite GT label are retained. The candidate with the smallest Euclidean distance to the anchor in the PCA feature space is then selected as the informative opposite sample. If no candidates satisfy these criteria, an anchor-opposite sample is randomly selected from the original set.

Specifically, the opposite-class samples are adaptively selected based on the updated feature space, where informative opposite-class samples are identified for each anchor. This epoch wise updating mechanism continuously refreshes the constructed triplets and prevents easy negatives from dominating the training process, thereby maintaining effective metric-learning supervision and avoiding premature saturation of the triplet loss.

\textbf{Base supervision:}
In this framework, the total loss combines focal loss and triplet loss to provide complementary supervision. Focal loss addresses class imbalance for classification, while triplet loss enhances feature discrimination by improving intra-class compactness and inter-class separability. 

The focal loss is formulated as:
\begin{equation}
	\mathcal{L}{focal}(p_t) = - \alpha_t (1 - p_t)^\gamma \log(p_t)
\end{equation}
where $p_t$ is the model that predicts the probability that the target belongs to the true category (\textit{p: the sample is in the same category, 1 - p: the sample is in the opposite category}). The $\alpha_t$ weighting factors of category, the value is computed as the proportion of each class in the training dataset. The $\gamma$ is a tunable focusing parameter for dynamic adjustment of loss weights for samples with different levels of difficulty.

The total loss is formulated as:
\begin{equation}
	\mathcal{L}_{total}=\mathcal{L}_{focal}+\lambda\mathcal{L}_{triplet}
\end{equation}
where $\lambda$ balances the classification and metric-learning objectives. Unless otherwise specified, $\lambda=1$ is adopted to provide equal contribution from both objectives. The influence of different $\lambda$ values is evaluated in Table~\ref{ab_tw}. 

During optimization, focal loss is calculated only on the anchor sample, while the triplet loss utilizes all three samples to optimize feature distribution. Since anchors are randomly selected from both categories during training, every sample receives classification supervision. Therefore, simultaneously optimization enables the shared backbone to simultaneously learn classification-aware and metric-aware representations.

\subsection{Implementation details} \label{3.4}	
The proposed RapidMitDet is implemented based on the PyTorch framework with NVIDIA GeForce RTX 3090 GPU. Our detection model is optimized using stochastic gradient descent (SGD) with a weight decay of 5e-4, utilizing 640$\times$640 pixels patches with a fixed bounding box size of 50$\times$50 pixels for preprocessing of training data.  For classification, we employ the Adam optimizer with the same weight decay and a learning rate of 1e-4, where the center point of the detected bounding box from the first stage is aligned with the ground-truth annotation, and the corresponding region is cropped into 64$\times$64 pixels patches for binary classification (mitosis versus non-mitosis). To enhance data diversity, real-time augmentation is applied during training. Specifically, random resized cropping was performed with a scale range of [0.8, 1.0] and an aspect ratio range of [0.9, 1.1]. Random affine transformation was applied with rotation of $\pm 10^\circ$, translation of 0.05, and scaling range of [0.9, 1.1]. Horizontal and vertical flipping were applied with a probability of 0.5. Gaussian blur was randomly applied with probability 0.3 using a kernel size of 3 and $\sigma$ ranging from 0.1 to 2.0. Random sharpening was applied with probability 0.3 using a sharpness factor of 2.

\section{Experiments} \label{material}
In this section, we introduce the details of the public datasets and evaluation metrics in Sections~\ref{4.1} and~\ref{sec:metric}, respectively. We then conduct state-of-the-art (SOTA) comparisons, internal cross-validation, and computational complexity analysis to demonstrate the effectiveness and efficiency of RapidMitDet in Sections~\ref{SOTA},~\ref{internal}, and~\ref{complexity}. Furthermore, the clinical applicability of RapidMitDet is evaluated through external clinical validation in Section~\ref{external}. For methods evaluated under identical experimental settings, we cite the performance reported in the original publications. For methods without available results under our settings, we reproduced these methods by strictly following the training configurations described in the respective papers. Finally, the effectiveness of the proposed magnification alignment (MAG) and adaptive triplet learning (ATL) modules is further validated through ablation studies in Section~\ref{ablation}.
\subsection{Datasets} \label{4.1}
\textbf{TUPAC16-auxiliary:} The auxiliary dataset of the 2016 TUmor Prolif-eration Assessment Challenge (TUPAC16)~\citep{veta2019predicting} is composed of 73 breast cancer whole-slide images from three different institutions.The first 23 cases are the dataset that was previously released as part of the Assessment of Mitosis Detection Algorithms 2013  challenge (AMIDA13)~\citep{veta2015assessment}. The remaining 50 cases are from two different pathology centers, Each case is represented with one image region with area of 2 $mm^{2}$.

\textbf{MIDOG2021:} The MIDOG2021~\citep{aubreville2023mitosis} challenge dataset provides 1721 mitotic figures and 2714 non-mitotic figures in 200 selected region of interest (ROI) level images. Among them, the first 150 cases are annotated, and the last 50 cases are unannotated. And mitotic and non-mitotic figures do not exist uniformly in every image. 

\textbf{MIDOG2022:} The MIDOG2022~\citep{aubreville2024domain} challenge dataset provides 9501 mitotic figures and 2714 non-mitotic figures which are look-alike mitotic figures that were discarded by the consensus of the three pathologists in 354 labeled images. The challenge dataset annotated histologic tumor images from 7 different domains.

\textbf{MIDOG++:} The MIDOG++ dataset~\citep{aubreville2023comprehensive} represents an extension of the MIDOG 2021 and 2022 challenge datasets. It provides region of interest~(ROI) images from 503 histological specimens of 7 different tumor types. The total mitotic labels figure is 11,937.

Regarding the relationship among the MIDOG benchmarks, they are hierarchically constructed, with later benchmarks incorporating data from previous challenges. All experiments were independently conducted following their corresponding protocols. TUPAC16-auxiliary was additionally included as an independent benchmark.

\subsection{Evaluation metric}~\label{sec:metric}
In this work, we adopt the ${F}_1$ score as the primary evaluation metric, consistent with existing methods. The ${F}_1$ score represents a harmonic mean of precision and recall, providing a balanced assessment of detection performance. A detected object is classified as a true positive (TP) if the Euclidean distance between its predicted center and the ground truth center is less than 7.5 $\mu m$. Given that all images were acquired at 40× magnification (pixel size = 0.25 $\mu m$), this threshold corresponds to 30 pixels (7.5 $\mu m$ / 0.25 $\mu m$ per pixel). Detections exceeding this distance from any ground truth annotation are counted as false positives (FP), whereas undetected ground truth objects within this radius are recorded as false negatives (FN). The overall performance is evaluated by computing the ${F}_1$ score on all ${N}$ processed slides, defined as:

\begin{equation}
	F_1=\frac{2\sum_k^N\mathrm{TP}_k}{2\sum_k^N\mathrm{TP}_k+\sum_k^N\mathrm{FN}_k+\sum_k^N\mathrm{FP}_k}
\end{equation}

where $TP_{k}$ , $FN_{k}$ , and $FP_{k}$ are the number of true positive, false negative, and false positive detections on slide ${k}$ and ${N}$ is the total number of slides~\citep{aubreville2023mitosis, aubreville2024domain}. For matching, we used a KD-tree with a maximum Euclidean distance threshold of 7.5 \textmu{m} between detection and ground truth bounding box centroids to prevent duplicate counting of true positives (TPs) and exclude false positives (FPs) without valid correspondences.

For computational efficiency evaluation, the inference time was measured under two settings. Tables 1–3 report image-level inference time for individual benchmark images, whereas Table 4 reports WSI-level end-to-end inference time, including WSI loading, preprocessing, inference, post-processing, and result generation. All methods were evaluated under the same hardware and software conditions.

\begin{table*}[htb]
	\centering
	\setlength{\tabcolsep}{2.5pt}
	\caption{Comparison of our RapidMitDet with the existing SOTA method on TUPAC16-auxiliary, MIDOG2021 and MIDOG++ datasets. All methods were evaluated under identical runtime environments at 40$\times$ magnification, with (20$\times$) indicating performance tested at 20$\times$ magnification. Speed baseline is MDFS inference time at 40$\times$ magnification. (40$\times$ represents both stages at 40$\times$; 20$\times$ represents stage 1 at 20$\times$ followed by stage 2 at 40$\times$). F1 scores were adopted from the original publications, whereas inference times were measured using our in-house implementations.}
	\label{tid}
	\begin{tabular}{lllccc}
		\toprule
		Datasets & Methods & Details & F1 & Time & Speed \\ \toprule
		
		\multirow{6}{*}{\parbox{1.2cm}{\centering TUPAC16\\(auxiliary)}} & SK-UNet~\citep{wang2023generalizable} & Sk-UNet with Fourier Segmentation & 0.745 & - & - \\
		& MitDet~\citep{lin2025rethinking} & H-channel + ResNet Classification & 0.746 & 15.8s & 0.20$\times$ \\
		& UTS~\citep{fernandez2024uninformed} & UTS Teacher-Student distill Segmentation & 0.767 & - & - \\
		& MDFS~\citep{jahanifar2024mitosis} & EUNet Segmentation + EfficientNetB7 Classification & 0.767 & 3.16s & 1.00$\times$ \\
		& RapidMitDet & Hyper-YOLO Detection + ResNet152 Classification & \textbf{0.849} & 1.07s & 2.95$\times$ \\
		& RapidMitDet (20$\times$) & Hyper-YOLO-Align Detection + ResNet152 Classification & 0.817 & \textbf{0.45s} & \textbf{7.02$\times$} \\ \hline
		
		\multirow{9}{*}{MIDOG21} & UTS~\citep{fernandez2024uninformed} & UTS Teacher-Student distill Segmentation & 0.713 & - & - \\
		& RetinaNet~\citep{wilm2021domain} & RetinaNet One Stage Detection & 0.720 & 1.72s & 2.42$\times$ \\
		& MitDet~\citep{lin2025rethinking} & H-channel + ResNet Classification & 0.789 & 21.4s & 0.19$\times$ \\
		& SK-UNet~\citep{wang2023generalizable} & Sk-UNet with Fourier Segmentation & 0.793 & - & -\\
		& BIA+PMS~\citep{li2022domain} & Box Instance-Aware and Pseudo-Mask Segmentation & 0.797 &  3.05s & 1.36$\times$\\
		& MDFS~\citep{jahanifar2024mitosis} & EUNet Segmentation + EfficientNetB7 Classification & 0.837 & 4.16s & 1.00$\times$ \\
		& Mask-RCNN~\citep{fick2021domain} & Mask-RCNN Segmentation + ResNet50 and DenseNet201 & 0.848 & 4.50s & 0.92$\times$\\
		& RapidMitDet (40$\times$) & Hyper-YOLO Detection + ResNet152 Classification  & \textbf{0.848} & 1.52s & 2.73$\times$ \\
		& RapidMitDet (20$\times$) & Hyper-YOLO-Align Detection + ResNet152 Classification & 0.826 & \textbf{0.74s} & \textbf{5.62$\times$}\\ \hline
		\multirow{6}{*}{MIDOG++} & TCS~\citep{kotte2022deep} & DETR Detection + EfficientNet-B7 Classification & 0.760 & - & - \\
		& USZ~\citep{lafarge2022fine} & ResNet70 with Sampling Hard-Negatives Classification & 0.790 & - & - \\
		& MDFS~\citep{jahanifar2024mitosis} & EUNet Segmentation + EfficientNet-B7 Classification & 0.800 & 5.16s & 1.00$\times$\\
		& OMG-Net~\citep{shen2024deep} & SAM Segmentation + ResNet18 Classification Ensemble & 0.850 & 23.0s & 0.22$\times$ \\
		& RapidMitDet (40$\times$) & Hyper-YOLO Detection + ResNet152 Classification & \textbf{0.852} & 2.38s & 2.17$\times$ \\
		& RapidMitDet (20$\times$) & Hyper-YOLO-Align Detection + ResNet152 Classification & 0.833 & \textbf{1.55s} & \textbf{3.33$\times$} \\ \bottomrule	
	\end{tabular}
\end{table*}

\subsection{Comparison with state-of-the-art methods}~\label{SOTA}
In this study, we conduct a comprehensive evaluation of our RapidMitDet model at 40× and 20× magnifications, benchmarking against recent state-of-the-art (SOTA) methods on the public datasets. The compared approaches include MDFS~\citep{jahanifar2024mitosis}, MitDet~\citep{lin2025rethinking}, UTS~\citep{fernandez2024uninformed}, BIA+PMS~\citep{li2022domain}, Mask R-CNN~\citep{fick2021domain}, RetinaNet~\citep{wilm2021domain}, OMG-Net~\citep{shen2024deep}, USZ~\citep{lafarge2022fine}, TCS~\citep{kotte2022deep}, and SK-UNet~\citep{wang2023generalizable}. Specifically, for TUPAC16-auxiliary and MIDOG21, the same train/validation/test splits as UTS were used. For MIDOG++, we followed the train/validation/test partition strategy reported in OMG-Net. Our method is evaluated in two distinct configurations: RapidMitDet (40×): Detection and classification both performed at 40× magnification; RapidMitDet (20×): Detection at 20× magnification followed by classification at 40× magnification.  

\hyperref[tid]{Table~\ref*{tid}} presents a quantitative comparison between our RapidMitDet framework and recent state-of-the-art methods across all three benchmark datasets. On the MIDOG21 dataset, our RapidMitDet achieves the SOTA F1-score performance at 40$\times$ magnification. Notably, our RapidMitDet~(20$\times$) maintains competitive accuracy with only a marginal reduction of 0.02 F1-score compared to RapidMitDet~(40$\times$), while delivering a remarkable 50\% improvement in computational efficiency. Impressively, our model outperforms numerous specialized methods designed specifically for 40$\times$ detection. This advantage remains consistent across two datasets, as evidenced by performance metrics on MIDOG++ that mirror our MIDOG21 findings. It shows that our method achieves an optimal trade-off between performance and efficiency and outperforms existing approaches.
\begin{figure*}[t]
	\centering
	\includegraphics[width=0.985\linewidth]{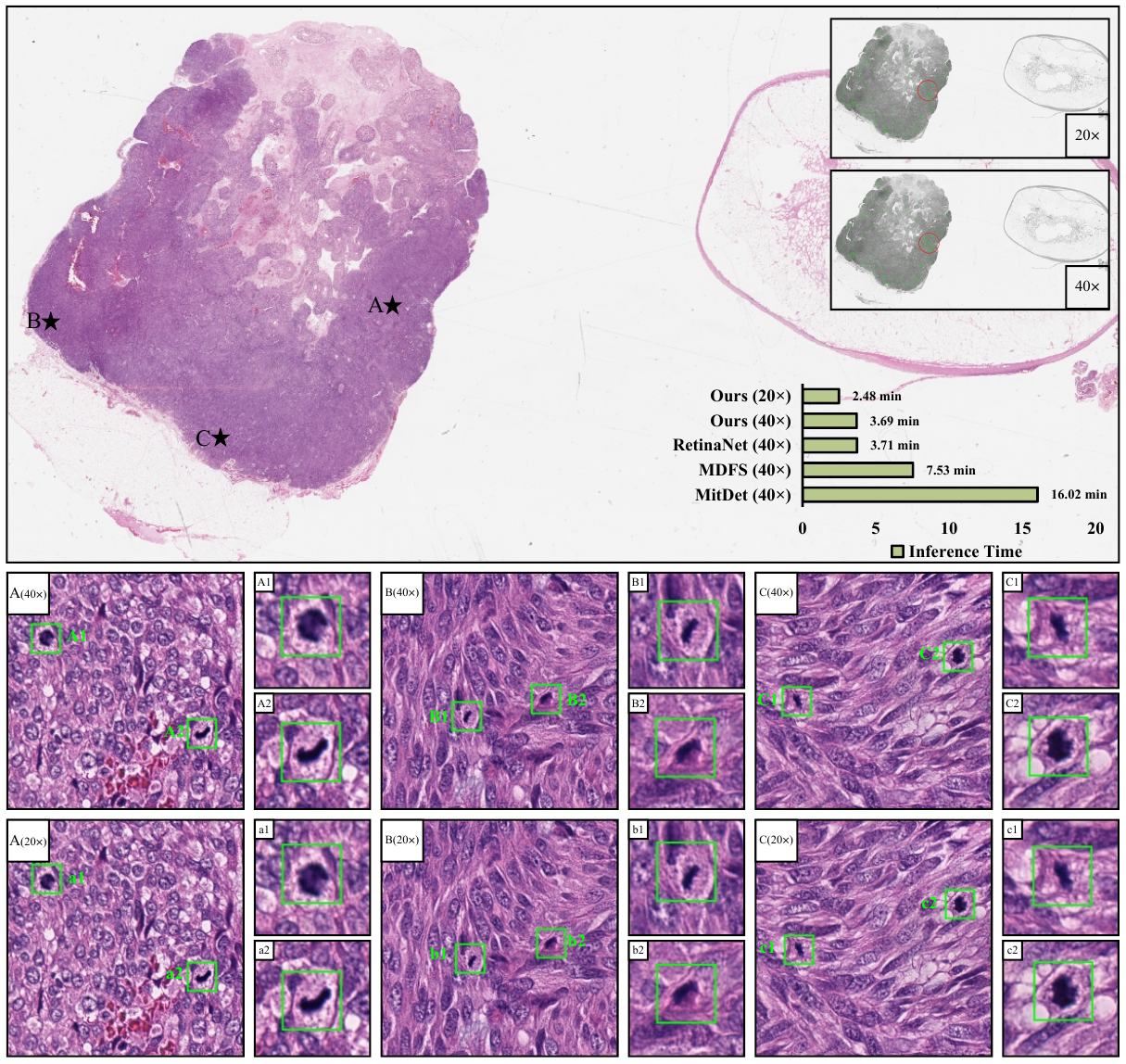}
	\caption{Inference comparison results for whole slide images (WSIs) from our private datasets are presented. In the upper right corner, masked thumbnails display the inference outcomes of our proposed model at both 40$\times$ and 20$\times$ magnification. The bar chart presents a comparative analysis of inference time between our proposed model and the existing state-of-the-art (SOTA) models. We further analyze three representative regions (A, B, and C) by magnifying them for detailed visualization.}
	\label{fig3}
\end{figure*}
\subsection{Internal Cross-validation}~\label{internal}
To further evaluate the reliability of our proposed model, we performed internal cross-validation on MIDOG21 and MIDOG22 datasets using identical 3-fold validation settings of MDFS. While the experimental results of MDFS are directly quoted from the source literature~\citep{jahanifar2024mitosis}, timing metrics (including runtime and speed) are obtained by reproducing their method in identical hardware configurations.

\hyperref[t3fold]{Table~\ref*{t3fold}} demonstrates the quantitative results of all. We bold the top approaches for every score. Our average F1 scores and average recall scores at 40$\times$ magnification achieved the best score. Meanwhile, our model demonstrates nearly equivalent performance at 20$\times$ magnification compared to the 40$\times$ setting. At 20$\times$ magnification setting, our model not only demonstrates a performance improvement 14\% compared to the previous SOTA MDFS but also reduces inference time by 59\%. To further validate the robustness and reliability of our model, we conducted identical experiments on the MIDOG22 dataset. The results also demonstrate that our model achieves comparable performance at 20× magnification as our model at 40$\times$, and simultaneously improves computational efficiency by 69\% compared to MDFS at 40$\times$. Overall, our proposed model has been shown to facilitate rapid inference at lower magnification, with performance that is closely aligned with that of inference at higher magnification.

\begin{table*}[htbp]
	\caption{Results of internal 3-fold cross-validation experiments on MIDOG21 and MIDOG22 datasets. The reported time is the average ROI image processing duration measured in seconds. Speed baseline is MDFS inference time at 40$\times$ magnification. (40$\times$ represents both stages at 40$\times$; 20$\times$ represents stage 1 at 20$\times$ followed by stage 2 at 40$\times$). MDFS metrics were adopted from the original publications, whereas inference times were measured using our in-house implementations.}
	\centering
	\setlength{\tabcolsep}{4pt}
	\label{t3fold}
	\begin{tabular}{lcccccccc}
		\toprule
		Dataset & Magnification & Methods & Setting & F1 & Precision & Recall & Time & Speed \\ \toprule
		\multirow{11}{*}{MIDOG21} & \multirow{6}{*}{40$\times$} & \makecell[c]{MDFS \\ \citep{jahanifar2024mitosis}} & Mean ± SD & 0.785 ± 0.004 & \textbf{0.801 ± 0.021} & 0.771 ± 0.020 & 4.16s & 1.00$\times$ \\ \cline{3-9}
		& & \multirow{4}{*}{\makecell[c]{Ours}} & Fold1  & 0.805 & 0.801 & 0.809 & \multirow{4}{*}{1.52s} & \multirow{4}{*}{2.73$\times$}\\
		& & & Fold2  & 0.774 & 0.780 & 0.769 &  \\
		& & & Fold3  & 0.789 & 0.788 & 0.791 &  \\ 
		& & & Mean ± SD  & \textbf{0.789 ± 0.015} & 0.790 ± 0.033 & \textbf{0.790 ± 0.008} & \\ \cline{2-9}
		
		& \multirow{5}{*}{20$\times$} & \makecell[c]{MDFS \\ \citep{jahanifar2024mitosis}} & Mean ± SD & 0.768 ± 0.011 & 0.794 ± 0.035 & 0.744 ± 0.049 & 1.82s & 2.29$\times$\\ \cline{3-9}
		& &\multirow{4}{*}{\makecell[c]{Ours}} & Fold1  & 0.803 & 0.803 & 0.803 & \multirow{4}{*}{\textbf{0.74s}} &\multirow{4}{*}{\textbf{5.62$\times$}}\\
		& & & Fold2  & 0.771 & 0.777 & 0.762 &  \\
		& & & Fold3  & 0.773 & 0.785 & 0.762 &  \\ 
		& & & Mean ± SD  & 0.782 ±  0.023 & 0.788 ±  0.015 & 0.776 ±  0.031 &  \\ \cline{1-9}
		
		\multirow{11}{*}{MIDOG22} & \multirow{6}{*}{40$\times$} & \makecell[c]{MDFS \\ \citep{jahanifar2024mitosis}} & - & 0.816 & \textbf{0.831} & 0.801 & 4.52s & 1.00$\times$ \\ \cline{3-9}
		& & \multirow{4}{*}{\makecell[c]{Ours}} & Fold1  & 0.823 & 0.811 & 0.835 & \multirow{4}{*}{2.09s} & \multirow{4}{*}{2.16$\times$}\\
		& & & Fold2  & 0.824 & 0.816 & 0.831 &  \\
		& & & Fold3  & 0.816 & 0.826 & 0.806 &  \\ 
		& & & Mean ± SD  & \textbf{0.821 ± 0.005} & 0.817 ± 0.003 & \textbf{0.824 ± 0.018} & \\ \cline{2-9}
		
		& \multirow{4}{*}{20$\times$} & \multirow{4}{*}{\makecell[c]{Ours}} & Fold1  & 0.811 & 0.806 & 0.816 & \multirow{4}{*}{\textbf{1.39s}} &\multirow{4}{*}{\textbf{3.25$\times$}}\\
		& & & Fold2  & 0.823 & 0.826 & 0.821 &  \\
		& & & Fold3  & 0.810 & 0.823 & 0.796 &  \\ 
		& & & Mean ± SD  & 0.814 ±  0.009 & 0.818 ±  0.008 & 0.811 ±  0.010 &  \\ 
		
		\bottomrule
	\end{tabular}
\end{table*}

\subsection{Comparison with computational complexity}~\label{complexity}
We also measure the computational complexity of our model with SOTA models including MitDet, MDFS, BIA+PMS, and RetinaNet on MIDOG21 training set. In addition to measuring time and speed, we also used parameters (number of model parameters) and FLOPs~(floating-point operations per second) as metrics to evaluate the computational complexity of methods.

\hyperref[cc]{Table~\ref*{cc}} demonstrates that the inference time of our proposed model, which is achieved by 0.74 seconds, compared to MitDet, MDFS, BIA+PMS, and RetinaNet, our model demonstrates improvements of 97\%, 82\%, 76\%, and 57\% in inference efficiency, respectively. For parameters, our model uses 87M parameters vs MDFS 76M. However MDFS uses egmentation-based approach requires extra localization time, while our detection formulation enables faster coordinate transformation. Thus, with comparable parameters, our model can be a sped-up inference.

\begin{table*}[th]
	\centering
	\caption{Comparison with computational complexity of our RapidMitDet and SOTA models on MIDOG21 datasets. All methods were evaluated under identical runtime environments at 40$\times$ magnification, with (20$\times$) indicating performance tested at 20$\times$ magnification. Speed baseline is MDFS inference time at 40$\times$ magnification. (40$\times$ represents both stages at 40$\times$; 20$\times$ represents stage 1 at 20$\times$ followed by stage 2 at 40$\times$). Results obtained using our in-house reimplementations.}
	\label{cc}
		\begin{tabular}{lcccc}
			\toprule
			Methods & Parameters & FLOPs & Time~(seconds) & Speed\\ \toprule
			\makecell[l]{RetinaNet~\citep{wilm2021domain}} & 34M & 303G & 1.72 & 2.42$\times$\\
			\makecell[l]{BIA+PMS~\citep{li2022domain}} & 33M & 152G & 3.05 & 1.36$\times$\\
			\makecell[l]{MDFS~\citep{jahanifar2024mitosis}} & 76M & 48G & 4.16 & 1.00$\times$\\
			\makecell[l]{MitDet~\citep{lin2025rethinking}} & 105M & 1631G & 21.4 & 0.19$\times$\\
			RapidMitDet~(40$\times$) & \multirow{2}{*}{87M} & \multirow{2}{*}{122G} & 1.52 & 2.73$\times$\\
			RapidMitDet~(20$\times$) &  &  & \textbf{0.74} & \textbf{5.62$\times$}\\ \bottomrule
		\end{tabular}
\end{table*}
\subsection{Clinical external validation}~\label{external}

In this section, we conduct a clinical external validation comparing to demonstrate the potential for clinical application of our RapidMitDet on in-house whole slide images~(15 patients with 36 histopathological slides of gastrointestinal stromal tumors (GIST)) corrected from Southern Hospital, Southern Medical University.

In clinical practice, the mitotic score (MS) for GIST is determined by mitosis counting~(MC) within a standardized 5 $mm^{2}$ microscopic field, following the NCCN guidelines  ~\citep{nccn2026gist}, which recommend using a standardized 5 mm$mm^{2}$ counting area because the area represented by high-power fields varies across different microscope configurations, with subsequent classification into three grades according to predefined threshold:
\begin{equation}MS=
	\begin{cases}
		0,  MC<5 \\
		1,  5\leq MC\leq10 \\
		2,  MC>10  
	\end{cases}.
\end{equation}

For the clinical validation, reference mitotic scores were established through the consensus assessment of two experienced pathologists. The pathologists independently evaluated all WSIs under blinded conditions, with disagreements resolved through consensus discussion. Both the AI system and the pathologists assessed the same WSIs, and the final mitotic score for each assessment was determined by the highest-count hotspot. For statistical analysis, mitotic scores were aggregated at the patient level, and agreement between the AI-derived and consensus pathologist scores was quantified using the intraclass correlation coefficient (ICC). Furthermore, we quantified the WSI-level inference time of each method (RapidMitDet, MitDet, MDFS, and RetinaNet) to compare their computational efficiency.

\begin{table}[]
	\centering
	\caption{Results of WSI-level external validation. All methods using MIDOG21-trained model on in-house dataset and evaluated under identical runtime environments at 40$\times$ magnification, with (20$\times$) indicating performance tested at 20$\times$ magnification. Speed baseline is MDFS inference time at 40$\times$ magnification. (40$\times$ represents both stages at 40$\times$; 20$\times$ represents stage 1 at 20$\times$ followed by stage 2 at 40x). Results obtained using our in-house reimplementations.} 
	\label{cwsi}
		\begin{tabular}{lcccc}
			\toprule
			Method & ICC[95\%CI] & Time & Speed\\ \toprule
			MitDet~\citep{lin2025rethinking} & 0.866 [0.60-0.96] & 31.9min & 0.37$\times$\\
			MDFS~\citep{jahanifar2024mitosis} &  0.732 [0.20-0.91] & 11.9min & 1.00$\times$\\
			RetinaNet~\citep{wilm2021domain} & 0.686 [0.02-0.89] &8.39min & 1.42$\times$\\
			Ours (40$\times$) & \textbf{0.879} [0.64-0.96]& 7.02min & 1.70$\times$\\
			Ours (20$\times$) & 0.822 [0.47-0.94]& \textbf{3.95min} & \textbf{3.01$\times$}\\ \bottomrule
		\end{tabular}
\end{table}

In \hyperref[cwsi]{Table~\ref{cwsi}}, quantitative results demonstrate that our model achieves an ICC score of 0.822 between 20$\times$ inference results and pathologists, indicating promising agreement and potential for clinical application. However, the limited sample size and single-institution setting introduce uncertainty, as reflected by the wide 95\% confidence interval [0.47, 0.94]. Further multicenter validation with larger cohorts is needed to establish its reliability and generalizability. Our method requires only 3.95 minutes per WSI, outperforming RetinaNet (53\% faster), MDFS (67\% faster) and MitDet (88\% faster). For a fair comparison, all methods were trained on the MIDOG21 dataset. It should be noted that MDFS relies on pixel-level segmentation annotations, which may limit its applicability when such annotations are unavailable in clinical scenarios. Our method shows promising clinical applicability in real-world workflows. As illustrated in \hyperref[fig3]{Fig.~\ref{fig3}}, we present the visualization of one representative WSI. The inference results of RapidMitDet at 20$\times$ and 40$\times$ magnifications show remarkable consistency, indicating that the proposed magnification alignment effectively improves cross-magnification feature consistency.

\subsection{Ablation studies}~\label{ablation}
In this work, our proposed RapidMitDet incorporates two key innovations: the magnification alignment (MAG) module and adaptive triplet learning (ATL). To demonstrate their individual contributions, we perform comprehensive ablation studies using internal validation sets from the MIDOG21 and MIDOG22 datasets, with quantitative results presented in \hyperref[tATL]{Table~\ref*{tATL}}.

For the MIDOG21 dataset, rows 1-2 in the table present ablation study on ATL. We can observe that under 40$\times$ magnification, RapidMitDet with ATL achieves an improvement in F1-score of over 0.02 compared to the baseline without ATL, with consistent improvements observed across all evaluation metrics. Rows 3-4 present the ablation study of MAG, demonstrating that incorporating MAG improves all performance metrics at 20$\times$ magnification. Notably, the computational savings primarily arise from performing inference at 20$\times$ rather than 40$\times$, while MAG helps mitigate the performance degradation associated with lower-magnification inference through cross-magnification feature alignment. The MAG module can also minimize missed detections and false positives in initial screening, optimizing the followed fine-grained classification stage. Most notably, the synergistic combination of both MAG and ATL modules yields superior performance compared to using either module alone, which demonstrates their complementary effects in simultaneously enhancing both model efficiency and detection accuracy.
\begin{table*}[t] 
	\centering
	\caption{Ablation study on magnification alignment~(MAG) and adaptive triplet learning~(ATL) on MIDOG21 and MIDOG22 datasets. Speed baseline is our model inference time at 40$\times$ magnification. (40$\times$ represents both stages at 40$\times$; 20$\times$ represents stage 1 at 20$\times$ followed by stage 2 at 40$\times$)}
	\label{tATL}
	\begin{tabular}{lcccccccc}
		\toprule
		Datasets & Magnification & MAG & ATL & F1 & Precision & Recall & Time~(seconds) & Speed \\ \toprule
		\multirow{4}{*}{MIDOG21} & \multirow{2}{*}{40$\times$} & - & - & 0.795 & 0.800 & 0.789 & \multirow{2}{*}{1.52} & \multirow{2}{*}{1.00$\times$}\\
		& & - & \ding{51} & 0.805 & 0.801 & 0.809 & & \\ \cline{2-9}
		& \multirow{3}{*}{20$\times$} & - & - & 0.773 & 0.785 & 0.762 & \multirow{3}{*}{0.74} & \multirow{3}{*}{2.05$\times$}\\
		& & \ding{51} & - & 0.787 & 0.794 & 0.780 & & \\
		& & \ding{51} & \ding{51} & 0.801 & 0.797 & 0.804 & & \\  \hline
		\multirow{4}{*}{MIDOG22} & \multirow{2}{*}{40$\times$} & - & - & 0.803 & 0.845 & 0.765 & \multirow{2}{*}{2.17} & \multirow{2}{*}{1.00$\times$}\\
		& & - & \ding{51} & 0.823 & 0.811 & 0.835 & & \\ \cline{2-9}
		& \multirow{3}{*}{20$\times$} & - & - & 0.775 & 0.823 & 0.732 & \multirow{3}{*}{1.39} & \multirow{3}{*}{1.56$\times$}\\
		& & \ding{51} & - & 0.801 & 0.824 & 0.778 & & \\
		& & \ding{51} & \ding{51} & 0.823 & 0.826 & 0.821 & & \\ \bottomrule
	\end{tabular}
\end{table*}

\begin{table}[t]
	\centering
	\caption{Effect of incorporating detector-generated false-positive samples into the second-stage classifier training on MIDOG21 datasets at 40$\times$ magnification.}
	\label{ab_fp}
	\begin{tabular}{cccccc}
		\hline
		Detector-generate FPs & F1 & Precision & Recall & \\
		\hline
		w  & 0.805 & 0.801 & 0.809  \\
		w/o  & 0.773 & 0.799 & 0.745  \\
		\hline
	\end{tabular}
\end{table}

\begin{table}[t]
	\centering
	\caption{Comparison of different bounding box annotation strategies for mitosis detection on MIDOG21 datasets at 40$\times$ magnification.}
	\label{ab_box}
	\begin{tabular}{cccccc}
		\hline
		Annotation strategy & F1 & Precision & Recall & \\
		\hline
		Official 50×50 boxes  & 0.805 & 0.801 & 0.809  \\
		Hover-Net adaptive boxes  & 0.736 & 0.721 & 0.752  \\
		\hline
	\end{tabular}
\end{table}

\begin{table}[t]
	\centering
	\caption{Comparison of different weighting factor ($\lambda$) of the triplet loss on MIDOG21 datasets at 40$\times$ magnification. The $\lambda$ controls the relative contribution of the triplet loss in Eq. (6).}
	\label{ab_tw}
	\begin{tabular}{cccccc}
		\hline
		$\lambda$ & F1 & Precision & Recall & \\
		\hline
		0.1  & 0.738 & 0.820 & 0.671  \\
		0.5  & 0.715 & 0.825 & 0.631  \\
		1.0  & 0.805 & 0.801 & 0.809  \\
		2.0  & 0.796 & 0.793 & 0.799  \\
		\hline
	\end{tabular}
\end{table}

\begin{figure*}[t]
	\centering
	\includegraphics[width=0.985\linewidth]{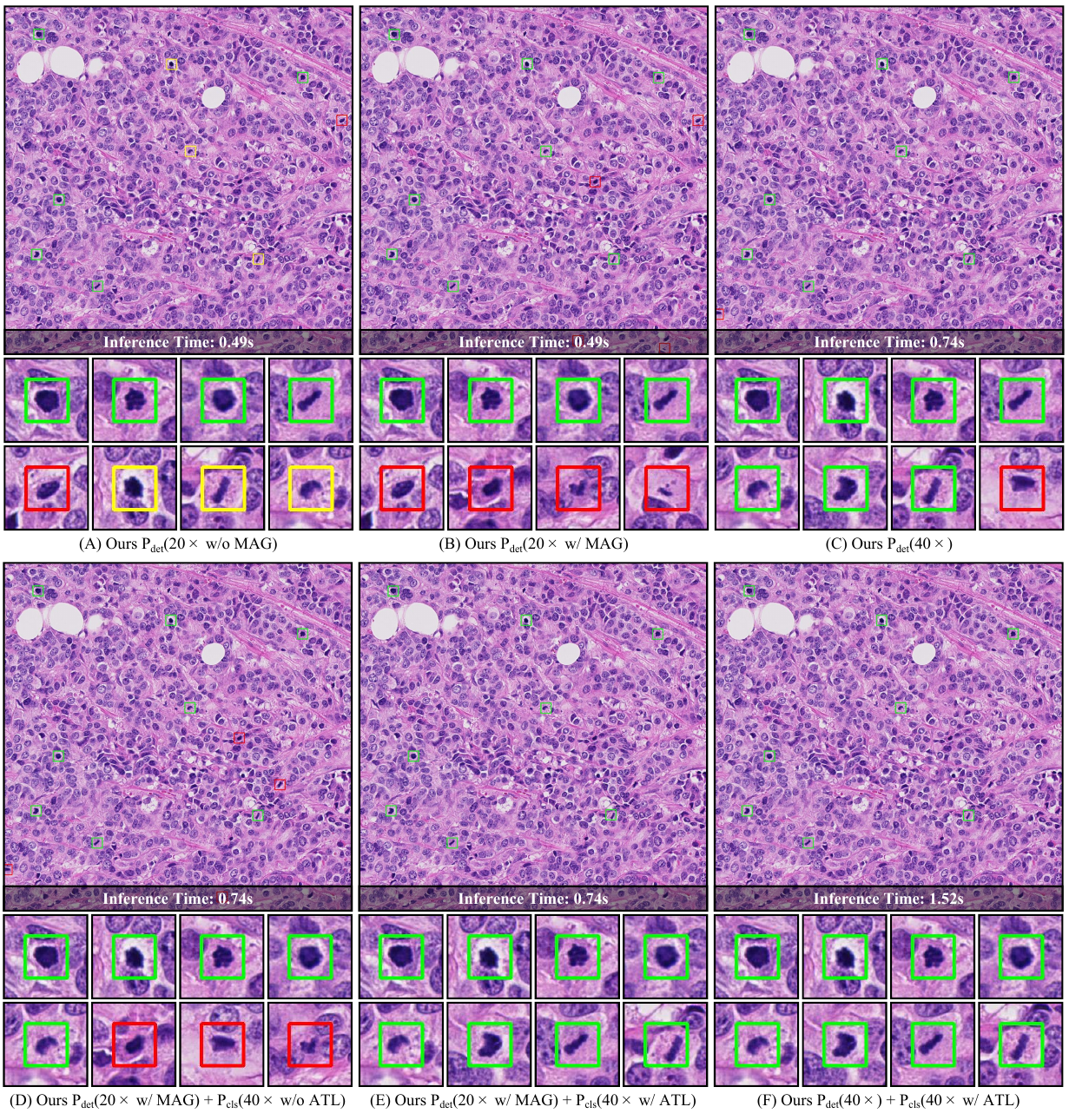}
	\caption{figure}{Qualitative results of ablation studies on our proposed magnification alignment~(MAG) module and adaptive triplet learning~(ATL) module. We visualize the zoom-in patches of the detection results ($\text{P}_\text{det}$: Phase 1 detection; $\text{P}_\text{cls}$: Phase 2 classification. Rectangle border color green:true positive, red: false positive, yellow: false negative). Image is collected from the MIDOG21 dataset. }
	\label{fig4}
\end{figure*} 
\hyperref[fig4]{Fig.\ref*{fig4}} illustrates the incremental improvements achieved by integrating the MAG and ATL modules. Panel (A) displays inference results without MAG, revealing significant false negatives. The introduction of MAG in (B) substantially reduces these false negatives, yielding markedly improved sensitivity. However, a performance gap persists compared to the 40$\times$ baseline in (C). This gap is effectively bridged in (E) through ATL integration, which not only matches the accuracy of the top-performing model (F) but also improves the inference efficiency by 51\%. Collectively, these results validate that each proposed module independently contributes to the improved performance of mitosis detection.

To evaluate performance consistency across domains, MIDOG2022 was divided into seven domains based on tissue type and scanner characteristics. 10 images from each domain were randomly selected for testing, with the rest used for training. As shown in \hyperref[fig-domain]{Fig.\ref*{fig-domain}}, Vahadane-based~\citep{vahadane2016structure} stain augmentation consistently improves F1-scores across all domains, demonstrating enhanced robustness to staining variations. The mitotic figure distribution across domains is also provided. Domains E and F contain more mitotic figures, which may partially contribute to their higher performance, while domain-level variations may also arise from differences in tissue characteristics and scanner conditions.

\begin{figure*}[tb]
	\centering
	\includegraphics[width=\linewidth]{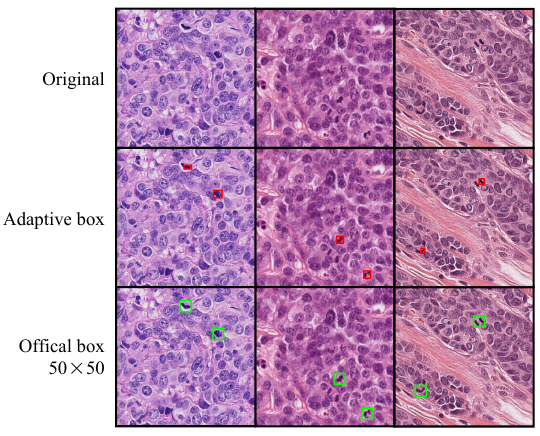}
	\caption{Comparison between the official MIDOG21 annotations and segmentation-derived adaptive bounding boxes. top: original histopathology patch, center: nucleus segmentation predicted by Hover-Net, bottom: official 50×50 bounding-box annotation. }
	\label{fig-box}
\end{figure*}

\begin{figure*}[tb]
	\centering
	\includegraphics[width=\linewidth]{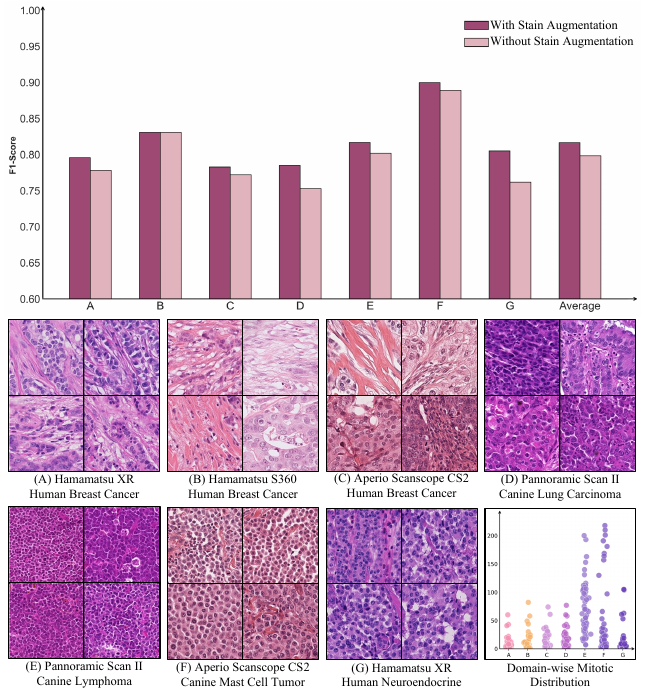}
	\caption{Qualitative results of ablation studies on MIDOG22: F1-score comparison across domains (tissue types and scanners).} 
	\label{fig-domain}
\end{figure*}

\section{Discussion} \label{discussion}

\subsection{Generalizability of RapidMitDet}

RapidMitDet demonstrates promising generalizability across multiple public mitosis detection benchmarks with diverse tissue types and imaging conditions. This robustness can be attributed to the complementary effects of magnification alignment and adaptive triplet learning, where the former alleviates magnification-related distribution shifts and the latter enhances the discriminative capability of learned representations.

\subsection{Efficiency of RapidMitDet}

A major advantage of RapidMitDet is its improved computational efficiency for large-scale digital pathology analysis. By introducing MAG, the proposed method reduces the dependence on high-magnification inference while preserving discriminative information required for mitotic identification. The image-level efficiency evaluation on public benchmarks and the WSI-level end-to-end validation on clinical whole-slide images demonstrate that RapidMitDet achieves a favorable balance between accuracy and computational cost.

This improved efficiency is particularly important for clinical applications, where whole-slide images typically contain gigapixel-level information and require substantial computational resources. By reducing inference time while maintaining reliable mitotic assessment, RapidMitDet has the potential to facilitate practical deployment in routine digital pathology workflows.

\subsection{Limitations and Future Work}

Despite the promising results, RapidMitDet has several limitations. As a coarse-to-fine framework, its performance depends on the quality of the first-stage detector, where false positives increase computational burden and missed candidates cannot be recovered. Moreover, incorporating detector-generated false positives as hard negatives (non-mitosis) may introduce potential annotation noise due to incomplete annotations in public datasets. Although this strategy improves overall classification performance (Table~\ref{ab_fp}), future work will investigate more robust sample selection and noise-aware learning strategies to further reduce the impact of potential label noise.

Furthermore, conventional YOLO-based detectors typically use adaptive bounding boxes, whereas public mitosis benchmarks provide only fixed-size annotations. To investigate the impact of annotation strategies, we generated adaptive bounding boxes by combining Hover-Net~\citep{graham2019hover} segmentation masks with the original annotations, as shown in Fig.~\ref{fig-box}. However, segmentation errors may introduce localization uncertainty, resulting in lower performance than the official 50$\times$50 annotations (Table~\ref{ab_box}). Future work will explore more accurate instance segmentation methods and pixel-level annotated datasets to improve adaptive annotation generation.

Although external clinical validation demonstrated the potential for clinical application of RapidMitDet, it was conducted on a limited single-institution cohort (15 patients and 36 WSIs). Therefore, larger multi-center studies are required to further validate its reliability and generalizability in clinical practice.

\section*{Acknowledgments}
This work was funded National Natural Science Foundation of China (No.62622313, No.82372044, No.82572321); the National Science Foundation for Young Scientists of China (No.82502464); the Guangdong Basic and Applied Basic Research Foundation (No.2023B1515020043); the China Postdoctoral Science Foundation(No.2025M772908).

\appendix
\newpage

\bibliography{ref}
\end{document}